\documentclass[12pt]{article}

\begin{document}

\begin{titlepage}

\title{Asymptotic behavior and critical coupling in scalar
Yukawa model}

\author{  {\bf V. E. Rochev}\footnote{E-mail address: rochev@ihep.ru}
\\  State Research Center of the Russian Federation\\ 
 ``Institute for High Energy Physics''\\
of National Research Centre  ``Kurchatov Institute``,\\ Protvino, Russia}
\date{}

\end{titlepage}
\maketitle
\begin{abstract} 
 The solution of the equation for the phion propagator in the leading order
of the $1/N$ -- expansion for a vector-matrix model 
with  interaction $g(\phi_a)^*\phi_b\chi_{ab}$ in four dimensions
shows 
a change of the asymptotic behavior 
in the deep Euclidean region in a vicinity of
 a certain critical value of the coupling constant.
\end{abstract}

In this report I will talk about one critical phenomenon found in a four-dimensional model.

I consider a vector-matrix model of  the complex scalar field $\phi_a$ (phion)
  and real scalar mass-less field $\chi_{ab}$ (chion)
with  interaction $g(\phi_a)^*\phi_b\chi_{ab}$ in four dimensions
($a, b=1,\cdots , N$).
  This model, known
as well as a scalar Yukawa model, is used in nuclear
physics as a simplified version of the Yukawa model without spin
degrees of freedom, as well as an effective model of
the interaction of scalar quarks (squarks).
  As
the simplest model of the interaction of fields,  this model often used
as a prototype of more substantive theories to elaborate
the  various non-perturbative approaches in
the quantum field theory.

 The solution of the equation for the phion propagator in the leading order
of the $1/N$ -- expansion
shows 
a change of the asymptotic behavior 
in the deep Euclidean region in a vicinity of
 a certain critical value of the coupling constant.
For small values of the coupling the
propagator behaves as free, which is consistent with the
wide-spread opinion  about the dominance of  perturbation theory
for this super-renormalizable model. In the strong-coupling 
region, however, the asymptotic behavior changes 
dramatically -- the propagator in the deep Euclidean region
tend to some constant limit.
 
The similar change of asymptotic behavior was found also with solution
of the system of Schwinger--Dyson equations in so-called two-particle
approximation \cite{Rochev13}.

The existence  of a critical coupling constant in the
scalar Yukawa model was noticed by practically all authors
that
have investigated this model using different methods.
This critical constant is generally regarded as
a limit on the coupling constant
for a self-consistent
description of the model by some method.
In our approach, however, the self-consistent solution
for propagator exists also in the strong-coupling region,
and the existence of the critical coupling
  looks more like as a phase
transition in accordance
with the general definition of the phase transition as a sharp
change of properties of the model with a smooth change of parameters.

 The Lagrangian
of the model is
\begin{equation}
   {\cal L}= -\partial_\mu(\phi_a)^*\partial_\mu\phi_a-m^2(\phi_a)^*\phi_a-
\frac{1}{2}(\partial_\mu\chi_{ab})^2+\frac{g}{\sqrt{N}}(\phi_a)^*\phi_b\chi_{ab}
\label{Lagr}
\end{equation}
($a, b=1,\cdots , N$). 

The technique of construction of the $1/N$ -- expansion for such models is well-known
 (\cite{tHooft, Slavnov}), and the leading-order equation for the 
 phion
propagator  $\Delta_{ab}(x)=\delta_{ab}\Delta(x)$ is
\begin{equation}
\Delta^{-1}(x)=(m^2-\partial^2)\delta(x)
-g^2D_c(x)\Delta(x).
\label{Delta0}
\end{equation}
Here $x\in E_4$ and $D_c=-1/\partial^2$.

At $N=1$ this equation coincides with the leading-order equation of the ladder
expansion (see \cite{Rochev15, Rochev16}), and
 in the language of Feynman diagrams it corresponds
 to the summation of ladder graphs.

To eliminate ultraviolet divergences in  equation (\ref{Delta0}) 
it is necessary to introduce counter-terms of phion-field renormalization
   and mass.
The normalization of the renormalized propagator $\Delta (p^2) $
at zero momentum
leads to the  renormalized equation in momentum space
\begin{equation}
\Delta^{-1}(p^2)=m^2+p^2+\Sigma_r(p^2),
\label{Delta}
\end{equation}
where $\Sigma_r(p^2)=\Sigma(p^2)-\Sigma(0)-p^2\Sigma'(0)$ is the renormalized
mass operator, $m$ is the renormalized mass, and 
$$
\Sigma(p^2)=-g^2\int \frac{d^4q}{(2\pi)^4}\;D_c(p-q)\Delta(q).
$$
After angle integrations we obtain the integral equation:
\begin{equation}
\Delta^{-1}(p^2)=m^2+(1-\lambda)p^2 +
2\lambda m^2 \int_{0}^{p^2} \Delta(q^2)\Big(1-\frac{q^2}{p^2}\Big) dq^2
\label{Delta_int}
\end{equation}
where
$$
\lambda\equiv\frac{g^2}{32\pi^2m^2}
$$
-- dimensionless coupling.

This equation can be reduced to the non-linear differential equation
\begin{equation}
 \frac{d^2}{(dp^2)^2}\big(p^2\Delta^{-1}(p^2)\big)=2(1-\lambda) + 2\lambda m^2\Delta(p^2).
\label{Delta_int_diff}
\end{equation}

We shall look for the positive solutions ($\Delta^{-1}(p)>0$)
 of the equation for the propagator in the region of large momenta.
Negative solutions necessarily contain Landau singularities and are therefore physically unacceptable.

In the weak-coupling region $\lambda<1$ 
the asymptotic solution at large $p^2$ is
$$
\Delta(p)= \frac{1}{(1-\lambda)p^2}\bigg(1-\frac{2\lambda m^2}{(1-\lambda)^2}\,
\frac{\ln p^2/m^2}{p^2}+O(1/p^2)\bigg)
$$
This asymptotic solution  is self-consistent in
 region 
$\lambda<1$ and corresponds to the asymptotically-free behavior of propagator.

At $\lambda=1$ the equation for the propagator can be reduced to singular Emden-Fowler equation:
$$
\ddot{y} = 2t y^{-1}.
$$
Here 
$$
y=\frac{p^2}{(m^2)^2}\Delta^{-1}(p),\;\; t=\frac{p^2}{m^2}.
$$
According to this, the asymptotic behavior of  propagators in the critical point
$\lambda=1$
has the form
$$ 
\Delta(p)=\sqrt{\frac{3}{8 m^2 p^2}}\,\bigg(1+O(1/p^2)\bigg)
$$
at large $p^2$ and drastically differs from the  asymptotically-free behavior in
the weak-coupling region.

In the strong-coupling region $\lambda>1$,  the  equation for the propagator has
the positive approximate solution
\begin{equation}
\Delta^{-1}(p^2)\simeq \frac{\lambda m^2}{\lambda-1}\bigg[1-\frac{1}{\lambda-1}\,
\sqrt{\frac{m^2}{2\lambda p^2}}\;J_1\Big(2(\lambda-1)\sqrt{\frac{2p^2}{\lambda m^2}}\Big)\bigg]
\label{StrongProp}
\end{equation}
(here $J_1$ is the Bessel function), which has the right normalization in zero
($
\Delta^{-1}(0)=m^2
$)
and the constant asymptotic behavior
$$
\Delta(\infty)=\frac{\lambda-1}{\lambda m^2}.
$$.

At $\lambda=1$  the asymptotic behavior of the propogator ($1/p$) is a medium among
the free behavior     $1/p^2$ at $\lambda<1$ 
and the constant--type behavior in strong--coupling region  $\lambda>1$.
A sharp change of  asymptotic behavior in the vicinity of the critical
value is a behavior that is characteristic for a phase transition.
This phase transition is not associated with a symmetry breaking, and in 
this sense is similar on the phase transition  "gas--liquid".
The weak coupling region can be roughly classified as the gaseous phase and
the strong coupling region (where  a kind of
  localization of correlators exists, see below) -- to the liquid.
This analogy, of course, is a quite shallow. A type and characteristics of this phase transition
  can be defined as the result of a more detailed  study.

This phase transition is similiar, in a sense, to the phenomenon of re-arrangement of physical
vacuum in the strong external field (see \cite{Greiner} and refs. therein).

The phion
propagator
in the strong-coupling region asymptotically approaches to a constant. It 
is not something unexpected, if we remember the well-known  conception
of the static ultra-local approximation.
In this  approximation,  all the Green functions are combinations of $\delta$-functions
in the coordinate space
 that are constants in momentum space.
Of course, this approximation is very  difficult for a physical 
interpretation.
Nevertheless, it can be considered as a starting point for
 an expansion in inverse powers of the coupling constant, i.e.,
as a leading approximation of the strong-coupling expansion  (see, e.g., \cite{Rivers}).

Note, that in  contrast to the ultra-local approximation, our solution are
free from the interpretation problem, since for the small
momenta it  has the quite traditional pole behavior.

Strong-coupling propagator (\ref{StrongProp}) has very interesting shell
structure in the four-dimensional Euclidean x--space. Doing Fourier transform of (\ref{StrongProp})
we obtain:
\begin{equation}
\Delta^{-1}(x)=\int \frac{d^4 p}{(2\pi)^4}\,e^{-ipx}\Delta^{-1}(p) =
\frac{\lambda m^2}{\lambda-1}\bigg(\delta^4(x)-\frac{m^2}{8\pi^2 (\lambda-1)^2}\,\delta(x^2-x^2_0)\bigg),
\end{equation}
where the quantity
$$
x_0=(\lambda-1)\sqrt{\frac{8}{\lambda m^2}} \sim \lambda_{compton}
$$
can be considered as the size of the phion localization region (a ``radius'' of the phion ). 

This radius increases as the coupling increases, which is very natural for the strong--coupling regime.

 $\bullet$ The main result   is finding of 
 the  change of  asymptotic behavior
in the scalar Yukawa model.
   The equation for the phion propagator in the leading order of $1/N$--expansion
has   self-consistent positive
solutions in the Euclidean region not only in the weak-coupling
region, (where  a dominance of the perturbation theory in this model is obvious),
but also in the strong-coupling region.
 Of  interest is the further study of this critical phenomenon, 
in particular,
the investigations  of amplitudes in this model and the search for
 analogs in other, more realistic models.

\end{document}